\documentstyle[aps,eqsecnum]{revtex}

\begin{document}
\draft
\title
{\bf 
Energy level statistics in weakly disordered systems: \\
from quantum to diffusive regime
}
\author{Naohiro Mae and Shinji Iida}
\address{
Faculty of Science and Technology, Ryukoku University, Otsu 520-2194, Japan
}
\date{January 21, 2003}
\maketitle

\begin{abstract}
We calculate two-point energy level correlation function in weakly disorderd 
metallic grain with taking account of localization corrections to the 
universal random matrix result. 
Using supersymmetric nonlinear $\sigma$ model and exactly integrating out 
spatially homogeneous modes, we derive the expression valid for arbitrary 
energy differences from quantum to diffusive regime for the system with broken 
time reversal symmetry. 
Our result coincides with the one obtained by Andreev and Altshuler 
[Phys. Rev. Lett. {\bf 72}, 902 (1995)] where homogeneous modes are 
perturbatively treated. 
\end{abstract}

\pacs{PACS number(s): 05.40.-a, 72.15.Rn, 05.45.Mt}

\section{Introduction}

Much research interest has been attracted by
the universal properties of quantum systems with randomness; 
the randomness arises from various sources such as 
complexity of the system  for the case of, e.g., complex nuclei, 
stochasticity for disorderd conductors and instability of 
classical trajectories for chaotic systems \cite{GMW}. 
If wave functions have enough time to diffuse throughout the system, 
the spatial strucure of the system becomes immaterial.    
For such cases, the system is called to belong to the ergodic 
regime where the statistical properties have been extensively studied. 
For example, correlation functions of the energy levels \cite{BH} or 
the scattering matrices \cite{HW,MI} are shown to have the universal forms 
where specific features of individual systems are represented 
by a few parameters \cite{GMW}.
These universal forms can be calculated with use of 
standard random matrix models belonging to pertinent Gaussian ensembles.

If the relevant time scale is comparable with the diffusion time,  
we need to turn to spatial dependent parts of the correlation functions 
which are not included in the universal random matrix results. 
These spatial dependent parts for weakly disordered systems are 
the subject of this paper.   
The weakly disorderd systems are characterized by the condition 
that $g \gg 1$ where $g$ is the dimensionless conductance.
The spatial dependent parts of the correlation functions are expected to appear
as higher order corrections with respect to $1/g$ 
to the universal random matrix results.
The two-point energy level correlation function
$R(s)$ (see Eq.~(\ref{Rs})) with corrections up to $1/g^2$ 
has been calculated by several authors:
In their seminal work, using diagrammatic perturbation theory, 
Altshuler and Shklovskii \cite{AS} obtained the smooth part of 
$R(s)$ valid for large energy differences $s \gg 1$.
Deviations from the universal statistics are determined by the 
diffusive classical dynamics of the system.
However, diagrammatic perturbation does not give an oscillatory part of $R(s)$ 
characteristic of the universal random matrix result.

The $1/g^2$ corrections including the oscillatory part were first obtained 
by Kravtsov and Mirlin (KM) \cite{KM} 
using the supersymmetric $\sigma$ model \cite{Ef}, 
the effective theory of a particle moving through a disorderd conductor.
Their result is, however, valid only for small energy differences $s \ll g$ 
\cite{PtoWD}. 
Later, also using the supersymmetric $\sigma$ model, 
Andreev and Altshuler (AA) \cite{AA}  
claimed that the oscillatory part can be obtained from 
perturbation around a novel saddle point 
in additon to the ordinary one.
Their result relies on perturbation and hence is restricted to the 
region of large energy differences $s \gg 1$.
The AA's result was subsequently reproduced with use of 
the replica \cite{KamMez} and Keldysh \cite{AK} $\sigma$ model.
In quantum chaotic systems, the analysis based on the semiclassical periodic 
orbit theory \cite{BK} and the ballistic $\sigma$ model \cite{AASA} 
led to similar results to the AA's.

In all these results, deviations from the universal form 
 is expressed in terms of eigenvalues of a certain operator 
or its spectral determinant. 
This operator is a diffusion operator for the case of 
disorderd conductors or a Perron-Frobenius operator for 
quantum chaotic systems \cite{AAA}.
We can surmise from these results that universal forms also exist 
in the deviations from universal random matrix resuls
where parameters representing system specific features
are now the spectra of this operator.

At present the available results are yet insufficient 
because KM's result is valid for $s \ll g$ and AA's one is valid for $s \gg 1$.
The method to calculate an expression valid for entire range of energy differences is desirable \cite{dffsv}.
This paper presents one such method. 
Extending the KM's procedure, 
we calculate $1/g^2$ corrections to 
the two-point energy level correlation function 
\begin{equation}
 R(s)=\Delta^{2}\langle \rho(E)\rho(E+\omega)\rangle \label{Rs}
\end{equation}
valid for arbitrary $\omega$, 
where $s=\omega/\Delta$, $\Delta$ is the mean level spacing, 
$\rho(E)$ is the density of states at energy $E$, 
and $\langle \ldots \rangle$ denotes 
averaging over realization of the random potential.
We consider disorderd systems with broken time reversal symmetry
because the pertinent random matrix ensemble (unitary ensemble)
is the simplest among the three classical ensembles
 (unitary, orthogonal, and symplectic). 
We follow the KM's procedure but do not use $\omega$ expansion which 
limits the range of their result's validity to small $\omega$.
As a result, we verify that the AA's expression gives the correct $g \gg 1$ 
asymptotic at arbitrary $\omega$ for unitary ensemble.
For the unitary case, invoking a mathematical theorem, 
it is implicitly shown \cite{Zirn} that 
 AA's result is exact in spite of their use of perturbation.
Here, we want to show it by explicit calculation which 
we expect to be  applicable in other ensembles with some modifications.

\section{Model and method} \label{sec: sec2}

The supersymmetric $\sigma$ model is a field theory 
whose variables $Q$ are supermatrices \cite{Ef}.  
Statistical properties of a particle moving through random potentials
can be derived from a generating functional of $Q$.
The expression for $R(s)$ reads
\begin{equation}
 R(s) = \frac{1}{16V^{2}} \mbox{Re} \int dQ\, e^{-S(Q)} 
 \left[ \int d\vec{r}\, \mbox{str}\, k \Lambda Q(\vec{r}) \right]^{2},\label{tpcf}
\end{equation}
\begin{equation}
 S(Q) = \frac{\pi}{4V}\int d\vec{r}\, \mbox{str}\, 
 \left[ \frac{D}{\Delta} (\nabla Q(\vec{r}))^{2} + 2\mbox{i} s^{+} \Lambda Q(\vec{r})\right]. 
 \label{action}
\end{equation}
Here $\mbox{str}$ denotes the supertrace, 
$Q(\vec{r}) = {T(\vec{r})}^{-1} \Lambda T(\vec{r})$ is 
a $4\times 4$ supermatrix with $T(\vec{r})$ belonging to 
the coset space $U(1,1|2)$, $\Lambda=\mbox{diag} (1,1,-1,-1)$, $k=\mbox{diag} (1,-1,1,-1)$, $V$ is the system volume, $D$ is the classical diffusion constant, 
$s^{+}=s + \mbox{i} 0^{+}$, and $0^{+}$ is positive infinitesimal. 
We use the notaions of Ref.~\cite{Ef} everywhere.

The ordinary saddle point is $Q = \Lambda$ and 
AA's novel one is $Q = - k \Lambda$.
The expansion around these saddle points and calculation of the functional 
integral by Gaussian approximation gives AA's result.
The quadratic term of the action $S(Q)$ defines the diffusion propagator 
in terms of which a pertubation series is constructed  by Wick's contraction. 
When $s$ goes to zero, the diffusion propagator diverges
for spatially homogeneous modes (which we hereafter call the zero-momentum modes, or zero modes for short, because their momentum is zero).
Thus their approach is restricted to the region $s \gg 1$.
  
KM have overcome the dfficulty of zero mode divergence 
by treating the zero modes and the other 
spatially inhomogeneous modes with non-zero momenta 
(heareafter called non-zero modes) separately.
The matrix $Q(\vec{r})$ is decomposed in the follwing way\cite{KM}: 
\begin{equation}
 Q(\vec{r}) = {T_{0}}^{-1} \tilde{Q}(\vec{r}) T_{0}, \label{decom}
\end{equation}
where $T_{0}$ is a spatially homogeneous matrix from the 
coset space and $\tilde{Q}(\vec{r})$ describes all non-zero modes 
(see Eqs.~(\ref{prtrb}), (\ref{W})). 
After perturbative calculation of the contributions by non-zero modes, 
there remains a defnite integral over a few zero mode variables, 
which can be evaluated without use of perturbation.
This amalgamation of  both  perturbative and non-perturbative calculation, 
in principle, could give the weak localization corrections 
valid for entire range of $s$.
In the course of calculation, however, assuming small $\omega$, 
KM expand the exponential of the energy term 
(the second term of the action (\ref{action})). 
Thus their result is restricted to $s \ll g$.

The difficulty arising when the energy term is kept on the shoulder of 
exponent is that the propagators of non-zero modes become dependent on
zero mode variables. 
Then after eliminating non-zero modes by perturbation,
the resulting zero mode integral contains infinite sums and products of 
non-zero mode propagators (see Eq.(\ref{D})) as an integrand.
Hence the zero mode integral, 
though  still being defnite over a few variables, does not seem feasible. 
In this paper, we show this is not the case:
by replacing the order of integration, it turns out that 
we can first integrate out the zero mode exactly.
Then integration over the non-zero modes by perturbation can be done 
without any trouble.

We resume a summary of notations: 
When $\Delta \ll E_{c}$ (where $E_{c}=D/L^{2}$ and $L$ is the system size), 
the matrix $\tilde{Q}(\vec{r})$ fluctuates only weakly around $\Lambda$. 
Then the fluctuations can be treated by perturbation with 
the expansion parameter $1/g$ (where $g=E_{c}/\Delta$). 
A convinient parametrization is 
\begin{equation}
 \tilde{Q}(\vec{r})=(1-W(\vec{r})/2) \Lambda (1-W(\vec{r})/2)^{-1},
 \label{prtrb}
\end{equation}
where 
\begin{equation}
 W(\vec{r}) = 
 \left(
 \begin{array}{cc}
  0 & W_{12}(\vec{r}) \\
  W_{21}(\vec{r}) & 0
 \end{array}
 \right) ,
 \label{W}
\end{equation}
$W_{21}(\vec{r}) = k{W_{12}(\vec{r})}^{\dagger}$.
$W_{12}(\vec{r})$ is expanded as $W_{12}(\vec{r})=\sum_{\vec{q}\neq 0}\phi_{\vec{q}}(\vec{r}) {W_{12}}_{\vec{q}}$ where $\phi_{\vec{q}}$ is an eigenfunction of the diffusion operator $-D \nabla^{2}$ with an eigenvalue $D {|\vec{q}|}^{2}$. 
$\{\phi_{\vec{q}}(\vec{r})\}$ constitute a complete orthonormal set and $\phi_{0}(\vec{r})=1/\sqrt{V}$. 
Thus 
$\int d\vec{r}\, W(\vec{r}) \propto \int d\vec{r}\, W(\vec{r}) \phi_{0}(\vec{r}) = 0$.
The Jacobian of the transformation Eqs.~(\ref{decom}) and (\ref{prtrb}) from the variable $Q$ to $\{T_{0}, W\}$  is 
\begin{equation}
 J\left(W\right)=1-\frac{1}{16 V^{2}}\int d\vec{r} d\vec{r'} 
 \left(\mbox{str}\, W_{12}(\vec{r}) W_{21}(\vec{r'})\right)^{2}+O(W^{5}). 
 \label{Jacob}
\end{equation}
The derivation is given in Appendix A.

The spatially homogeneous supermatrix $T_{0}$ is parametrized in a quasi-diagonalized form 
\begin{equation}
 T_{0}=U^{-1} \hat{T}_{0} U
\end{equation}
with `eigenvalue' matrix $\hat{T}_{0}$ given by 
\begin{equation}
 \hat{T}_{0}=
 \left(
 \begin{array}{cc}
   \cos\frac{\hat{\theta}}{2}
   & \mbox{i}\sin\frac{\hat{\theta}}{2} e^{\mbox{i} \hat{\varphi}}\\
   \mbox{i}\sin\frac{\hat{\theta}}{2} e^{-\mbox{i} \hat{\varphi}}
   & \cos\frac{\hat{\theta}}{2}
 \end{array}
 \right),
\end{equation}
\begin{equation}
 \hat{\theta}=
 \left(
 \begin{array}{cc}
   \theta_{F} & 0 \\
   0 & \mbox{i} \theta_{B}
 \end{array}
 \right),
\makebox[1cm]{}
 \hat{\varphi}=
 \left(
 \begin{array}{cc}
   \varphi_{F} & 0 \\
   0 & \varphi_{B}
 \end{array}
 \right),
\end{equation}
where $0 \leq \theta_{B} < \infty$, $0 \leq \theta_{F} \leq \pi$, 
$0 \leq \varphi_{B} \leq 2\pi$, $0 \leq \varphi_{F} \leq 2\pi$.
The `diagonalizing' matrix $U$ is given by 
\begin{equation}
 U=
 \left(
 \begin{array}{cc}
   v_{1} & 0 \\
   0 & v_{2}
 \end{array}
 \right),
\end{equation}
\begin{equation}
 v_{1}=
 \exp\left(
 \begin{array}{cc}
   0 & \xi_{1} \\
   -\xi_{1}^{*} & 0
 \end{array}
 \right),
\makebox[1cm]{}
 v_{2}=
 \exp \mbox{i} \left(
 \begin{array}{cc}
   0 & \xi_{2} \\
   -\xi_{2}^{*} & 0
 \end{array}
 \right),
\end{equation}
where $\xi_{1}$ and $\xi_{2}$ are anticommuting variables.
For this parametrization, the measure $dT_{0}$ is given by 
\begin{equation}
 dT_{0}=\frac{\sinh\theta_{B}\sin\theta_{F}}
             {\left(\cosh\theta_{B}-\cos\theta_{F}\right)^{2}}
        d\theta_{B} d\theta_{F} d\varphi_{B} d\varphi_{F} 
        d\xi_{1} d\xi_{1}^{*} d\xi_{2} d\xi_{2}^{*}.
\end{equation}

\section{Changing the variables}

To simplify the zero mode integration, we change the variables.
Substituting the decomposition (\ref{decom}) into Eq.~(\ref{tpcf}), 
we obtain 
\begin{equation}
 \mbox{str}\,\left( \nabla Q(\vec{r}) \right)^{2}
 =\mbox{str}\,\left( \nabla \tilde{Q}(\vec{r}) \right)^{2}, \label{Skinetic}
\end{equation}
\begin{equation}
 \int d\vec{r}\, \mbox{str}\, \Lambda Q(\vec{r}) 
 = \mbox{str}\, \Lambda {T_{0}}^{-1} Y T_{0}, 
 \label{Senergy}
\end{equation}
\begin{equation}
 \int d\vec{r}\, \mbox{str}\, k \Lambda Q(\vec{r}) = 
 \mbox{str}\, k \Lambda {T_{0}}^{-1} Y T_{0},
 \label{Ssource}
\end{equation}
where $Y = \int d\vec{r}\, \tilde{Q}(\vec{r})$. 
We make the supermatrix $Y$ a block diagonal form as follows (for detail see Appendix B):
\begin{equation}
 e^{X} Y e^{-X} = \hat{Q}, \label{BD}
\end{equation}
where $X$ is a block off-diagonal supermatrix and 
\begin{equation}
 \hat{Q}=\Lambda\left(V+\frac{1}{2}\int d\vec{r}\,W^{2}(\vec{r})
                     +\frac{1}{8}\int d\vec{r}\,W^{4}(\vec{r})
                     +O(W^{6})\right). \label{BDQ}
\end{equation}
Up to $1/g^{2}$ order calculated here, $e^{X} T_{0}$ can be replaced with $T_{0}$ (see Appendix B). Thus in Eqs.~(\ref{Senergy}) and (\ref{Ssource}), $Y$ 
can be replaced with $\hat{Q}$.

To eliminate the anticommuting variables of the zero modes in the action $S(Q)$, we change the variables as $U W U^{-1} \to W$. Eventually, we get 
\begin{equation}
 \int d\vec{r}\, \mbox{str}\, \Lambda Q(\vec{r}) 
 = \mbox{str}\, \Lambda {\hat{T}_{0}}^{-1} \hat{Q} \hat{T}_{0}, 
 \label{Senergy2}
\end{equation}
\begin{equation}
 \int d\vec{r}\, \mbox{str}\, k \Lambda Q(\vec{r}) = 
 \mbox{str}\, k \Lambda U^{-1} {\hat{T}_{0}}^{-1} \hat{Q} \hat{T}_{0} U. 
 \label{Ssource2}
\end{equation}

\section{Integration over the zero-momentum mode variables}


Expanding the pre-exponential term of Eq.~(\ref{tpcf}) in 
the anticommuting variables of the zero modes, we obtain 
\begin{equation}
 \left( \mbox{str}\, k \Lambda U^{-1} {\hat{T}_{0}}^{-1} \hat{Q} \hat{T}_{0} U \right)^{2}
 =
 \left( \mbox{str}\, k \Lambda {\hat{T}_{0}}^{-1} \hat{Q} \hat{T}_{0} \right)^{2}
 -
 2\xi_{1}^{*} \xi_{1} \xi_{2}^{*} \xi_{2} 
 \left(  \mbox{str}\, \Lambda {\hat{T}_{0}}^{-1} \hat{Q} \hat{T}_{0} \right)^{2}
 +
 \ldots \label{expndG}
\end{equation}
where we have used that $\mbox{str}\, \hat{Q}=0$ and the dots indicate terms which 
vanish after integration of the zero mode anticommuting variables.
We write the contribution by the first (second) term of Eq.~(\ref{expndG}) to 
$R(s)$ as $R_{1}(s)$ ($R_{2}(s)$).
With use of Parisi-Sourlas-Efetov-Wegner theorem \cite{PSEW}, 
integration of zero mode variables in $R_{1}(s)$ becomes 
the value of its integrand at $\theta_{B}=\theta_{F}=0$:
\begin{equation}
 R_{1}(s)=\frac{1}{16V^{2}} \mbox{Re} \int dW\, J\left(W \right) e^{-S(W)} 
          \left( \mbox{str}\, k \Lambda \hat{Q} \right)^{2} + O(1/g^{3}),
\end{equation}
 where
\begin{equation}
 S(W)=\frac{\pi}{4V}\mbox{str}\,\left[
      \frac{D}{\Delta}\int d\vec{r}\, \left( \nabla \tilde{Q}(\vec{r}) \right)^{2}
      +2\mbox{i} s^{+} \Lambda \hat{Q} \right].
\end{equation}
Integrating the zero mode anticommuting variables, we obtain 
\begin{equation}
 R_{2}(s) = 
 -\frac{1}{8 V^{2}} \mbox{Re} \int dW\, 
 d\theta_{B} d\theta_{F}\,
 J\left(W \right) 
 \frac{\sinh\theta_{B}\sin\theta_{F}}
      {\left(\cosh\theta_{B}-\cos\theta_{F}\right)^{2}} 
 \, e^{-S(\hat{T}_{0},W)} 
 \left( \mbox{str}\, \Lambda {\hat{T}_{0}}^{-1} \hat{Q} \hat{T}_{0} \right)^{2}
 +O(1/g^{3}),
\end{equation}
\begin{equation}
 S(\hat{T}_{0},W)=\frac{\pi}{4V}\mbox{str}\,\left[
      \frac{D}{\Delta}\int d\vec{r}\, 
      \left( \nabla \tilde{Q}(\vec{r}) \right)^{2}
      +2\mbox{i} s^{+} \Lambda {\hat{T}_{0}}^{-1} \hat{Q} \hat{T}_{0} \right].
\end{equation}
Here, 
\begin{equation}
 \mbox{str}\, \Lambda {\hat{T}_{0}}^{-1} \hat{Q} \hat{T}_{0} = 
 \cosh\theta_{B}\, \mbox{str}\, P(B) \Lambda \hat{Q}
 +\cos\theta_{F}\, \mbox{str}\, P(F) \Lambda \hat{Q},
\end{equation}
 where $P(B)=({\bf 1}-k)/2$ and $P(F)=({\bf 1}+k)/2$ and 
${\bf 1}$ is an unit matrix. 
 Using 
 $\cosh\theta_{B} = \lambda_{B}$, $\cos\theta_{F} = \lambda_{F}$, 
 $\mbox{str}\, P(B) \Lambda \hat{Q} = -2V\left(1+A\right)$, and  
 $\mbox{str}\, P(F) \Lambda \hat{Q} = 2V\left(1+B\right)$, 
 we obtain 
\begin{equation}
 R_{2}(s)=\frac{1}{2} \mbox{Re} \int dW\, J\left(W \right)
          \exp\left[-\frac{\pi D}{4V\Delta} \int d\vec{r}\, \mbox{str}\, 
                    \left( \nabla \tilde{Q}(\vec{r}) \right)^{2} \right]
          I(s,W) + O(1/g^{3}), 
\end{equation}
where
\begin{equation}
 I(s,W) = 
 \int_{1}^{\infty} d\lambda_{B} \int_{-1}^{1} d\lambda_{F} 
 f(\lambda_{B},\lambda_{F}) g(\lambda_{B},\lambda_{F}),
 \label{cmvint}
\end{equation}
\begin{equation}
 f(\lambda_{B},\lambda_{F}) =
 \frac{1}{\left(\lambda_{B}-\lambda_{F}\right)^{2}} 
 e^{ \mbox{i}\pi s^{+} \left( \lambda_{B}-\lambda_{F} \right) }, 
\end{equation}
\begin{equation}
 g(\lambda_{B},\lambda_{F}) =
 e^{ \mbox{i}\pi s^{+} \left( \lambda_{B}\,A-\lambda_{F}\,B \right) }
 \left[ \lambda_{B}\, \left(1+A\right)
        -\lambda_{F}\, \left(1+B\right) \right]^{2}.
\end{equation}


We introduce an indefinite integral of $f(\lambda_{B},\lambda_{F})$:
$
 \frac{\partial^{2}}{\partial \lambda_{F} \partial \lambda_{B}} 
 F(\lambda_{B},\lambda_{F})=f(\lambda_{B},\lambda_{F}),
$
\begin{eqnarray}
 F(\lambda_{B},\lambda_{F})&=&-\int_{-\infty}^{\lambda_{F}}db\,
 \int_{\lambda_{B}}^{\infty}da\,e^{\mbox{i}\pi s^{+} (a-b)}
 \frac{1}{(a-b)^{2}} \nonumber \\
 &=&-\int_{1}^{\infty} dx\, e^{\mbox{i}\pi s^{+} (\lambda_{B}-\lambda_{F})x}
 \left(\frac{1}{x}-\frac{1}{x^{2}}\right),
\end{eqnarray}
where we have used the transformation 
$(a-b)^{-2} = \int_{0}^{\infty} t e^{-t(a-b)} dt$ and 
$\mbox{i}\pi s^{+}-t=\mbox{i}\pi s^{+}x$.
Then partial integration gives 
\begin{eqnarray}
 I(s,W)&=&
 \left[\left[
 F(\lambda_{B},\lambda_{F}) g(\lambda_{B},\lambda_{F})
 \right]_{\lambda_{B}=1}^{\lambda_{B}=\infty}
 \right]_{\lambda_{F}=-1}^{\lambda_{F}=1}
 -\left[
 \int_{1}^{\infty} d\lambda_{B}\, F(\lambda_{B},\lambda_{F}) 
 \frac{\partial}{\partial \lambda_{B}} g(\lambda_{B},\lambda_{F})
 \right]_{\lambda_{F}=-1}^{\lambda_{F}=1} \nonumber \\
 &&-\left[
 \int_{-1}^{1} d\lambda_{F}\, F(\lambda_{B},\lambda_{F}) 
 \frac{\partial}{\partial \lambda_{F}} g(\lambda_{B},\lambda_{F})
 \right]_{\lambda_{B}=1}^{\lambda_{B}=\infty}
 +\int_{1}^{\infty} d\lambda_{B} \int_{-1}^{1} d\lambda_{F}
 F(\lambda_{B},\lambda_{F})
 \frac{\partial^{2}}{\partial\lambda_{B}\partial\lambda_{F}} 
 g(\lambda_{B},\lambda_{F}). \label{pardif}
\end{eqnarray}
After $\lambda$-integration, we obtain
\begin{eqnarray}
I(s,W)&=&
-\left[\left[
e^{ \mbox{i}\pi s^{+} \left( \lambda_{B}\,A-\lambda_{F}\,B \right) }
\int_{1}^{\infty} dx\, e^{\mbox{i}\pi s^{+} (\lambda_{B}-\lambda_{F})x}
{\cal I}(\lambda_{B},\lambda_{F},x)
\right]_{\lambda_{B}=1}^{\lambda_{B}=\infty}
\right]_{\lambda_{F}=-1}^{\lambda_{F}=1} \nonumber \\
&=&
 e^{ \mbox{i}\pi s^{+} \left( A-B \right) } \int_{1}^{\infty} dx\, {\cal I}(1,1,x)
-e^{ \mbox{i}\pi s^{+} \left( A+B \right) } 
\int_{1}^{\infty} dx\, e^{2\mbox{i}\pi s^{+} x} {\cal I}(1,-1,x), \label{II}
\end{eqnarray}
where
\begin{eqnarray}
{\cal I}(\lambda_{B},\lambda_{F},x)&=& 
\left(\frac{1}{x}-\frac{1}{x^{2}}\right)
\hat{A}^{-1}\hat{B}^{-1}
\left[ \lambda_{B}\, \left(1+A\right)
        -\lambda_{F}\, \left(1+B\right) \right]^{2} \nonumber \\
&&
-\frac{2}{\mbox{i}\pi s^{+}} \left(\frac{1}{x^{2}}-\frac{1}{x^{3}}\right)
\hat{A}^{-2}\hat{B}^{-2}
\left[ \hat{A}\left(1+B\right)
       +\hat{B}\left(1+A\right) \right]
\left[ \lambda_{B}\, \left(1+A\right)
        -\lambda_{F}\, \left(1+B\right) \right] \nonumber \\
&&
+\frac{2}{(\mbox{i}\pi s^{+})^{2}}\left(\frac{1}{x^{3}}-\frac{1}{x^{4}}\right)
\hat{A}^{-3}\hat{B}^{-3}
\left[ \hat{A}^{2}\left(1+B\right)^{2}
	   +\hat{A}\hat{B}
	    \left(1+A\right)\left(1+B\right)
	   +\hat{B}^{2}\left(1+A\right)^{2} \right]. 
\label{d2sI}
\end{eqnarray}
Here, $\hat{A} = 1+\frac{A}{x}$ 
and $\hat{B}= 1+\frac{B}{x}$. 
From Eq.~(\ref{II}) we can read that the propagator do not include 
zero mode variables.
Therefore the difficulty of KM's approach mentioned in Sec.~\ref{sec: sec2} is resolved. 
We write the contribution by the first (second) term of Eq.~(\ref{II}) to 
 $R_{2}(s)$ as $R_{2}^{(1,1)}(s)$ ($R_{2}^{(1,-1)}(s)$).

\section{Integration over the non-zero-momentum mode variables}

 In order to calculate the remaining integral over the non-zero mode variables $W$ with use of the Wick theorem, we need to know the contraction rules for the action 
\begin{equation}
 S_{0} = \frac{\pi}{4V}\int d\vec{r}\, \mbox{str}\, 
 \left[ 
 -\frac{D}{\Delta}\left(\nabla W(\vec{r})\right)^{2}
 + \mbox{i} s^{+} \left( \lambda_{B} P(B)+\lambda_{F} P(F) \right) W(\vec{r})^{2} 
 \right].
\end{equation}
The result is summarized as 
\begin{equation}
 \langle \mbox{str}\, {\cal X} W(\vec{r}) {\cal Y} W(\vec{r'}) \rangle 
 = - \frac{1}{2} \sum_{g,g'}\Pi(\vec{r},\vec{r'})_{g g'} 
 \left[
 \mbox{str}\, P(g) {\cal X} \; \mbox{str}\, P(g') {\cal Y} 
 - \mbox{str}\, \Lambda P(g) {\cal X} \; \mbox{str}\, \Lambda P(g') {\cal Y} 
 \right], 
\end{equation}
\begin{equation}
 \langle \mbox{str}\, {\cal X} W(\vec{r}) \; \mbox{str}\, {\cal Y} W(\vec{r'}) \rangle
 = - \frac{1}{2} \sum_{g, g'}\Pi(\vec{r},\vec{r'})_{g g'} 
 \left[ 
 \mbox{str}\, P(g) {\cal X} P(g') {\cal Y} 
 - \mbox{str}\, \Lambda P(g) {\cal X} \Lambda P(g') {\cal Y} 
 \right],
\end{equation}
 where ${\cal X}$ and ${\cal Y}$ are arbitrary $4\times 4$ supermatrices, $g$ and 
$g'$ are $B$ or $F$, 
\begin{equation}
 \langle \ldots \rangle = 
 {\cal D}^{-1}(s,\lambda_{B},\lambda_{F}) \int dW\, (\ldots) e^{-S_{0}},
\end{equation}
\begin{equation}
 {\cal D}(s,\lambda_{B},\lambda_{F}) = 
 \prod_{\vec{q} \ne \vec{0}}\frac{\Pi(q;\lambda_{B},\lambda_{B})
 \Pi(q;\lambda_{F},\lambda_{F})}
 {{\Pi(q;\lambda_{B},\lambda_{F})}^{2}},
\label{D}
\end{equation}
\begin{equation}
 \Pi(q; \lambda_{g},\lambda_{g'}) = \frac{2V}{\pi}
 \left( 
 \frac{D}{\Delta}q^{2}-\mbox{i} s^{+}\frac{\lambda_{g}+\lambda_{g'}}{2}
 \right)^{-1},
\end{equation}
\begin{equation}
 \Pi(\vec{r},\vec{r'})_{g g'}= \sum_{\vec{q} \ne \vec{0}} 
 \phi_{\vec{q}}(\vec{r}) \phi_{\vec{q}}^{*}(\vec{r'}) 
 \Pi(q; \lambda_{g},\lambda_{g'}).
\end{equation}


For $R_{1}(s)$, using the contraction rules with $\lambda_{B}=1$ and $\lambda_{F}=1$ we obtain
\begin{eqnarray}
 R_{1}(s)&=&
 \frac{1}{16V^{2}} \mbox{Re} 
 \left\langle 
 16V^{2}+\frac{1}{4}\left( 
 \int d\vec{r}\, \mbox{str}\, k{W(\vec{r})}^{2} \right)^{2} 
 \right\rangle_{(\lambda_{B},\lambda_{F})=(1,1)}
 +O(1/g^{3}) \nonumber \\
 &=&
 1+\frac{1}{8V^{2}} \mbox{Re} \sum_{\vec{q} \ne \vec{0}} 
 \Pi(q; 1,1)^{2}
 +O(1/g^{3}) \nonumber \\
 &=&
 1-\frac{1}{4\pi^{2}}\frac{d^{2}}{{ds}^{2}}\ln {\cal D}(s,1,-1) + O(1/g^{3}).
 \label{R1}
\end{eqnarray}


In order to get $R_{2}(s)$, we need to estimate the order of the each term of 
Eq.~(\ref{d2sI}) with use of the fact that 
(i) $W^{2} \sim O(1/g)$ for all $s$, 
(ii) the property of the exponential integral 
\begin{equation}
E_{n}(a) = \int_{1}^{\infty}\frac{e^{-a x}}{x^{n}} dx \label{Ei}
\end{equation}
where ${\rm Re}\, a > 0$ and $n \in {\bf N}$, and 
(iii) KM's result \cite{KM}: for $s < 1$,
\begin{equation}
R(s)=\sum_{n=0}^{\infty}\frac{C_{n}(s)}{g^{n}} \label{KMr}
\end{equation}
where $C_{n}(s)$ is $O(1)$.

\subsection{$(\lambda_{B},\lambda_{F})=(1,1)$ point}

Expanding ${\cal I}(1,1,x)$ in $W$, there appears 
a term proportional to $1/x$, $(A-B)^{2}/x$. 
The $x$-integration of this term, at first glance, seems to diverge.
Actually it causes no problem becaue it vanishes after $W$-integration. 
In order to obtain the corrections up to $1/g^{2}$, 
it is sufficient to expand the remaing terms up to the 4th order of $W$ 
for entire range of $s$ \cite{s};

\begin{eqnarray}
{\cal I}(1,1,x)&=& 
-\frac{1}{x^{2}}(A-B)^{2} \nonumber \\
&&
-\frac{2}{\mbox{i} \pi s^{+}}
\left[
2\left(\frac{1}{x^{2}}-\frac{1}{x^{3}}\right)(A-B)
+\left(\frac{1}{x^{2}}-\frac{4}{x^{3}}+\frac{3}{x^{4}}\right)(A^{2}-B^{2})
\right] \nonumber \\
&&
+\frac{2}{(\mbox{i} \pi s^{+})^{2}}
\left[
3\left(\frac{1}{x^{3}}-\frac{1}{x^{4}}\right)
+3\left(\frac{1}{x^{3}}-\frac{3}{x^{4}}+\frac{2}{x^{5}}\right)(A+B)
+\left(\frac{1}{x^{3}}-\frac{9}{x^{4}}+\frac{18}{x^{5}}-\frac{10}{x^{6}}\right)
(A^{2}+B^{2}+AB)
\right] \nonumber \\
&& +O(1/g^{3}).
\end{eqnarray}
After $x$-integration, we obtain 
\begin{equation}
\int_{1}^{\infty} {\cal I}(1,1,x) dx= 
-(A-B)^{2} 
-\frac{2}{\mbox{i} \pi s^{+}}(A-B) 
+\frac{1}{(\mbox{i} \pi s^{+})^{2}}
+O(1/g^{3}).
\end{equation}
After $W$-integration, we obtain
\begin{equation}
R_{2}^{(1,1)}(s)=-\frac{1}{2 \pi^{2} s^{2}}.
\label{R2p}
\end{equation}

\subsection{$(\lambda_{B},\lambda_{F})=(1,-1)$ point}

For $s \ge 1$, $E_{n}(-\mbox{i} 2\pi s^{+})$ in Eq.~(\ref{II}) is $O(1)$. 
Accordingly, by the same reason as the case of $(\lambda_{B},\lambda_{F})=(1,1)$ point, it is sufficient to expand ${\cal I}(1,-1,x)$ up to the 4th order of $W$ for entire range of $s$ again.
\begin{eqnarray}
{\cal I}(1,-1,x)&=& 
4\left[
\left(\frac{1}{x}-\frac{1}{x^{2}}\right)
+\frac{4}{\tilde{s}}\left(\frac{1}{x^{2}}-\frac{1}{x^{3}}\right)
+\frac{6}{{\tilde{s}}^{2}}\left(\frac{1}{x^{3}}-\frac{1}{x^{4}}\right)
\right]
\nonumber \\
&&
+4\left[
\left(\frac{1}{x}-\frac{2}{x^{2}}+\frac{1}{x^{3}}\right)
+\frac{2}{\tilde{s}}
\left(\frac{2}{x^{2}}-\frac{5}{x^{3}}+\frac{3}{x^{4}}\right)
+\frac{6}{{\tilde{s}}^{2}}
\left(\frac{1}{x^{3}}-\frac{3}{x^{4}}+\frac{2}{x^{5}}\right)
\right](A+B)
\nonumber \\
&&
+\left[
\left(\frac{1}{x}-\frac{5}{x^{2}}+\frac{7}{x^{3}}-\frac{3}{x^{4}}\right)
+\frac{4}{\tilde{s}}
\left(\frac{1}{x^{2}}-\frac{7}{x^{3}}+\frac{12}{x^{4}}-\frac{6}{x^{5}}\right)
+\frac{6}{{\tilde{s}}^{2}}
\left(\frac{1}{x^{3}}-\frac{9}{x^{4}}+\frac{18}{x^{5}}-\frac{10}{x^{6}}\right)
\right](A+B)^{2}
\nonumber \\
&&
+\left[
\left(\frac{1}{x^{3}}-\frac{1}{x^{4}}\right)
-\frac{4}{\tilde{s}}
\left(\frac{1}{x^{3}}-\frac{3}{x^{4}}+\frac{2}{x^{5}}\right)
+\frac{2}{{\tilde{s}}^{2}}
\left(\frac{1}{x^{3}}-\frac{9}{x^{4}}+\frac{18}{x^{5}}-\frac{10}{x^{6}}\right)
\right](A-B)^{2}
\nonumber \\
&&
+O(1/g^{3})
\end{eqnarray}
where $\tilde{s} = -\mbox{i} 2\pi s^{+}$.
Using the relation 
$
E_{n}(\tilde{s})+\frac{n}{\tilde{s}} E_{n+1}(\tilde{s})
 = e^{-\tilde{s}}/\tilde{s}
$, 
we obtain
\begin{equation}
\int_{1}^{\infty}dx\, e^{-\tilde{s}x}{\cal I}(1,-1,x)
=4\frac{e^{-\tilde{s}}}{{\tilde{s}}^{2}}
+\frac{1}{{\tilde{s}}^{2}}
\left[
e^{-\tilde{s}}
+2\int_{1}^{\infty}\,dx 
e^{-\tilde{s} x}\left(\frac{1}{x^{3}}-\frac{3}{x^{4}}\right)
\right](A-B)^{2}
+O(1/g^{3}),
\end{equation}
\begin{eqnarray}
e^{\mbox{i}\pi s^{+}(A+B)}\int_{1}^{\infty}dx\, e^{\mbox{i} 2\pi s^{+}x}{\cal I}(1,-1,x)
&=&
e^{\mbox{i}\pi s^{+}(A^{(2)}+B^{(2)})}
\frac{e^{\mbox{i} 2\pi s^{+}}}{(\mbox{i}\pi s^{+})^{2}} \nonumber \\
&&
+
e^{\mbox{i}\pi s^{+}(A^{(2)}+B^{(2)})}\left[
\frac{e^{\mbox{i} 2\pi s^{+}}}{\mbox{i}\pi s^{+}}(A^{(4)}+B^{(4)})
+
\frac{{\cal F}(s)}{4(\mbox{i}\pi s^{+})^{2}} (A^{(2)}-B^{(2)})^{2}
\right] \nonumber \\
&&
+O(1/g^{3}), \label{eqn1}
\end{eqnarray}
where 
$
{\cal F}(s)=
e^{\mbox{i} 2\pi s^{+}}
+2\int_{1}^{\infty}\,dx 
e^{\mbox{i} 2\pi s^{+} x}\left(\frac{1}{x^{3}}-\frac{3}{x^{4}}\right)
$.
The contribution by the second term of Eq.~(\ref{eqn1}) is $O(1/g^{3})$, 
because for $s \ge g$ it is obvious and for $s < g$ it is found after 
$W$-integration.

By the same way, 
the contributions by $O(W^{4})$ terms emerging from 
$\left( \nabla \tilde{Q}(\vec{r}) \right)^{2}$ and the jacobian $J(W)$ 
turn out to be $O(1/g^{3})$. 
Finally, we obtain  
\begin{eqnarray}
R_{2}^{(1,-1)}(s)
&=&
-\frac{1}{2} \mbox{Re} \int dW\, J\left(W \right) 
\exp\left[- \frac{\pi D}{4V\Delta} \int d\vec{r}\, \mbox{str}\, 
\left( \nabla \tilde{Q}(\vec{r}) \right)^{2} \right] 
e^{\mbox{i}\pi s^{+}(A^{(2)}+B^{(2)})} 
\frac{e^{\mbox{i} 2\pi s^{+}}}{(\mbox{i} \pi s^{+})^{2}}
+O(1/g^{3}) \nonumber \\
&=&
\frac{\cos 2\pi s}{2\pi^{2}s^{2}} {\cal D}(s,1,-1)
+O(1/g^{3}).
\label{R2n}
\end{eqnarray}

\section{Result and summary}
Equations~(\ref{R1}), (\ref{R2p}), and (\ref{R2n}) give 
\begin{eqnarray}
 R(s)
 &=&
 1-\frac{1}{4\pi^{2}}\frac{d^{2}}{{ds}^{2}}\ln {\cal D}(s,1,-1)
 -\frac{1}{2\pi^{2}s^{2}}
 +\frac{\cos 2\pi s}{2\pi^{2}s^{2}}{\cal D}(s,1,-1)+O(1/g^{3})
 \nonumber \\
 &=&
 1-\frac{1}{4\pi^{2}}\frac{d^{2}}{{ds}^{2}}\ln \tilde{\cal D}(s)
 +\frac{\cos 2\pi s}{2\pi^{2}} \tilde{\cal D}(s) +O(1/g^{3}),
\end{eqnarray}
 where $\tilde{\cal D}(s)= {\cal D}(s,1,-1)/s^{2}$ is the spectral determinant of the classical diffusion operator, in agreement with Refs.~\cite{AA,KamMez,AK}. Here the contribution of the $(\lambda_{B},\lambda_{F})=(1,1)$
\begin{equation}
-\frac{1}{4\pi^{2}}\frac{d^{2}}{{ds}^{2}}\ln \tilde{\cal D}(s)
=\frac{1}{8V^{2}}\mbox{Re}\sum_{\vec{q}}\Pi(\vec{q},1,1)^{2}
\end{equation}
is the ordinary perturbative result \cite{AS}.

In summary, 
using KM's separate treatment of the zero and non-zero modes, performing the 
integration over the zero modes exactly and subsequent 
integration over the non-zero modes perturbatively, 
we find the expression for the two-point energy level correlation function valid for arbitrary energy differences up to $1/g^{2}$ order for unitary ensemble. 
This expression is same as the one \cite{AA,KamMez,AK} obtained 
in a saddle-point approximation which in general valid only 
for $\omega \gg \Delta$ \cite{regulator}. 
By explicit calculation, we verify that 
the exactness of the saddle-point answer for unitary ensemble,
which is guaranteed in the ergodic regime due to the Duistermaat-Heckman 
theorem \cite{Zirn}, holds even in the diffusive regime. 
Since this is a specific feature of the unitary ensemble, it is very 
interesting to investigate the expressions for the other (orthogonal, 
symplectic, etc.) ones.

The reason why AA's novel saddle point appears is as follows: 
The zero-mode integral is carried out exactly.
Then the remaining expression for non-zero modes is evaluated at 
terminal points of the integral region, 
$1 \le \lambda_{B} < \infty , -1 \le \lambda_{F} < 1 $. 
These terminal points precisely correspond to 
the ordinary ($\lambda_{B}=\lambda_{F}=1$) 
and AA's novel saddle point ($\lambda_{B}=1, \lambda_{F}=-1$).
This correspondence between the terminal points of zero-mode variables 
and saddle points seems to hold for the orthogonal ensemble.
Hence, for the orthogonal case, a similar way of calculation may work: 
although exact integration of zero mode variables is no longer probable,  
the iteration of partial integration 
enables one to evaluate the integral on the terminal points up to 
a necessary order of $1/g$, if  
the additional terms generated by these iterations 
are only higher order term of $1/g$.
For the symplectic case, the correspondence seems more subtle because the novel saddle points are not isolated but consist of a saddle point maifold. 
Further investigation will be necessary for the symplectic case.

\section*{Acknowledgments}

We would like to thank K. Takahashi for valuable comments.
S.I. wish to acknowledge helpful discussions with participants 
in the workshop `` quantum chaos, theory and experiment  2002 '' 
held at Yukawa institute for theoretical physics, Kyoto, Japan, 
where part of this work was presented.

\appendix
\section{Calculation of the Jacobian}

We give the calculation of the Jacobian $J$ of the transformation $Q\to \{W, T_{0}\}$ because, to the best of our knowledge, there is no explicit expression of it in literature.
It is calculated by the following Gaussian integral : 
\begin{equation}
 \int d(\delta W) d(\delta T_{0}) \exp\left[
 - \int d\vec{r}\,\mbox{str}\, \left(\delta Q(\vec{r})\right)^{2}\right]
 =J^{-1}. \label{GIJ}
\end{equation}
Here $\delta Q(\vec{r})$ is variation of Eq.~(\ref{decom}) : 
\begin{equation}
 \delta Q(\vec{r})=
 T_{0}^{-1} \left( \delta \tilde{Q}(\vec{r})+
 \left[\tilde{Q}(\vec{r}), \delta T_{0}'\right] \right) T_{0},
\end{equation}
where $\delta T_{0}'= \delta T_{0}\, T_{0}^{-1}$. Then, 
\begin{equation}
 \int d\vec{r}\,\mbox{str}\, \left(\delta Q(\vec{r})\right)^{2}=
 \int d\vec{r}\,\mbox{str}\, \left(\delta \tilde{Q}(\vec{r})\right)^{2}
 +2\int d\vec{r}\,\mbox{str}\, \delta \tilde{Q}(\vec{r}) 
 \left[\tilde{Q}(\vec{r}), \delta T_{0}'\right]
 +\int d\vec{r}\, \mbox{str}\, \left[\tilde{Q}(\vec{r}), \delta T_{0}'\right]^{2}.
\end{equation}
The integrand of Eq.~(\ref{GIJ}) is expanded up to 4th order in $W$. 

The $\delta W$ dependent part of the integrand is 
\begin{eqnarray}
 & &
 \exp\left[
 - \int d\vec{r}\,\mbox{str}\, \left(\delta \tilde{Q}(\vec{r})\right)^{2}
 - 2\int d\vec{r}\,\mbox{str}\, \delta \tilde{Q}(\vec{r}) 
 \left[\tilde{Q}(\vec{r}), \delta T_{0}'\right]
 \right] \nonumber \\
 &=&
 \exp\left[
 \int d\vec{r}\, \mbox{str}\, (\delta W)^{2}
 \right] \nonumber \\
 &&
 \times \left\{ 1 
 +\frac{1}{2}\int d\vec{r}\,\mbox{str}\, W^{2} (\delta W)^{2}
 +2\left(\int d\vec{r}\,\mbox{str}\, X_{1}(\delta W)\delta T_{0}'
 \right)^{2} \right. \nonumber \\
 &&
 +2\left(\int d\vec{r}\,\mbox{str}\, X_{1}(\delta W)\delta T_{0}'\right)
 \left(\int d\vec{r}\,\mbox{str}\, X_{2}(\delta W)\delta T_{0}'\right) \nonumber \\
 &&
 +\frac{1}{8}\int d\vec{r}\,\mbox{str}\, W^{4}(\delta W)^{2}
 +\frac{1}{16}\int d\vec{r}\,\mbox{str}\, (W^{2}\delta W)^{2}
 +\frac{1}{8}\left[\int d\vec{r}\,\mbox{str}\, W^{2}(\delta W)^{2}\right]^{2}
 \nonumber \\
 &&
 +\left(\int d\vec{r}\,\mbox{str}\, X_{1}(\delta W)\delta T_{0}'\right)
 \left(\int d\vec{r}\,\mbox{str}\, X_{3}(\delta W)\delta T_{0}'\right) \nonumber \\
 &&
 +\frac{1}{2}
 \left(\int d\vec{r}\,\mbox{str}\, X_{2}(\delta W)\delta T_{0}'\right)^{2}
 +\frac{2}{3}
 \left(\int d\vec{r}\,\mbox{str}\, X_{1}(\delta W)\delta T_{0}'\right)^{4} \nonumber \\
 &&
 +\left.\left[\int d\vec{r}\,\mbox{str}\, W^{2} (\delta W)^{2}\right]
 \left(\int d\vec{r}\,\mbox{str}\, X_{1}(\delta W)\delta T_{0}'\right)^{2} 
 + O(W^{5}) \right\},
\end{eqnarray}
 where
\begin{equation}
 X_{n}(\delta W)= \sum_{k=0}^{n}(-1)^{k}W^{k} \delta W W^{n-k}.
\end{equation}
The $\delta W$ independent part is 
\begin{eqnarray}
 &&
 \exp\left(
 - \int d\vec{r}\, \mbox{str}\, \left[\tilde{Q}(\vec{r}), \delta T_{0}'\right]^{2}
 \right) \nonumber \\
 &=&
 \exp\left[ 2V\mbox{str}\, (\delta T_{0}'-\Lambda\delta T_{0}'\Lambda)\delta T_{0}'
 \right] \nonumber \\
 && 
 \times \left\{
 1-2\int d\vec{r}\, \mbox{str}\, \Lambda\delta T_{0}'\Lambda
 \left( W^{2}\delta T_{0}'-W\delta T_{0}'W\right) \right. \nonumber \\
 &&
 -\int d\vec{r}\, \mbox{str}\, \Lambda\delta T_{0}'\Lambda
 \left( W^{3}\delta T_{0}'-2W^{2}\delta T_{0}'W\right) \nonumber \\
 &&
 -\frac{1}{2}\int d\vec{r}\, \mbox{str}\, \Lambda\delta T_{0}'\Lambda
 \left( W^{4}\delta T_{0}'-2W^{3}\delta T_{0}'W+W^{2}\delta T_{0}'W^{2}\right) 
 \nonumber \\
 && \left.
 +2\left[\int d\vec{r}\, \mbox{str}\, \Lambda\delta T_{0}'\Lambda
 \left( W^{2}\delta T_{0}'-W\delta T_{0}'W\right) \right]^{2}
 +O(W^{5})\right\}. \label{dWind}
\end{eqnarray}
The integral of the $\delta W$ dependent part over $\delta W$ can be calculated by using the Wick theorem and the contraction rules:
\begin{equation}
 \langle 
 \mbox{str}\, {\cal X} \delta W(\vec{r}) {\cal Y} \delta W(\vec{r'}) 
 \rangle_{\delta W} 
 = \frac{1}{4} \left( \frac{1}{V} - \delta (\vec{r}-\vec{r'}) \right) 
 \left(
 \mbox{str}\, {\cal X} \; \mbox{str}\, {\cal Y} 
 - \mbox{str}\, \Lambda {\cal X} \; \mbox{str}\, \Lambda {\cal Y} 
 \right),
\end{equation}
\begin{equation}
 \langle 
 \mbox{str}\, {\cal X} \delta W(\vec{r}) 
 \; \mbox{str}\, {\cal Y} \delta W(\vec{r'}) 
 \rangle_{\delta W}
 = \frac{1}{4} \left( \frac{1}{V} - \delta (\vec{r}-\vec{r'}) \right) 
 \left( 
 \mbox{str}\, {\cal X} {\cal Y} - \mbox{str}\, \Lambda {\cal X} \Lambda {\cal Y} \right),
\end{equation}
where ${\cal X}$ and ${\cal Y}$ are arbitrary supermatrices and 
\begin{equation}
 \langle \ldots \rangle_{\delta W} = 
 \int d(\delta W)\, (\ldots) 
 \exp\left[\int d\vec{r}\, \mbox{str}\, (\delta W)^{2}\right] .
\end{equation}
The result is 
\begin{eqnarray}
 &&
 \int d(\delta W)\, \exp\left[ 
 - \int d\vec{r}\,\mbox{str}\, \left(\delta \tilde{Q}(\vec{r})\right)^{2}
 - 2\int d\vec{r}\,\mbox{str}\, \delta \tilde{Q}(\vec{r}) 
 \left[\tilde{Q}(\vec{r}), \delta T_{0}'\right] \right] \nonumber \\
 &=&
 1+\int d\vec{r}\, \mbox{str}\, \left(\delta T_{0}'+\Lambda\delta T_{0}'\Lambda\right)
 \left( W^{2}\delta T_{0}'-W\delta T_{0}'W\right) 
 +\int d\vec{r}\, \mbox{str}\, \Lambda\delta T_{0}'\Lambda
 \left( W^{3}\delta T_{0}'-2W^{2}\delta T_{0}'W\right) \nonumber \\
 &&
 -\frac{1}{8} \int d\vec{r}\,\mbox{str}\,\left(W^{2}\delta T_{0}'\right)^{2}
 +\frac{1}{8}\int d\vec{r}\, \mbox{str}\, \Lambda\delta T_{0}'\Lambda
 \left( 4W^{4}\delta T_{0}'-8W^{3}\delta T_{0}'W+5W^{2}\delta T_{0}'W^{2}\right) \nonumber \\
 &&
 +\frac{1}{2}\left[
 \int d\vec{r}\, \mbox{str}\, \left(\delta T_{0}'+\Lambda\delta T_{0}'\Lambda\right)
 \left( W^{2}\delta T_{0}'-W\delta T_{0}'W\right) 
 \right]^{2} \nonumber \\
 &&
 +\frac{1}{8V}\int d\vec{r}\,d\vec{r'}\, \mbox{str}\, 
 \left( \delta T_{0}'-\Lambda\delta T_{0}'\Lambda \right)
 \left( 2W(\vec{r})^{2}W(\vec{r'})^{2}\delta T_{0}'
 -2W(\vec{r})^{2}W(\vec{r'})\delta T_{0}'W(\vec{r'}) \right. \nonumber \\
 && \left.
 -2W(\vec{r})W(\vec{r'})^{2}\delta T_{0}'W(\vec{r'})
 +2W(\vec{r})^{2}\delta T_{0}'W(\vec{r'})^{2}
 +W(\vec{r})W(\vec{r'})\delta T_{0}'W(\vec{r'})W(\vec{r})\right)
 \nonumber \\
 &&
 -\frac{1}{64V} \int d\vec{r}\,\left(\mbox{str}\, W^{2}\right)^{2}
 +\frac{1}{64V^{2}} \left(\int d\vec{r}\,\mbox{str}\, W^{2}\right)^{2}
 +O(W^{5}).
\end{eqnarray}
Multiplying this by Eq.~(\ref{dWind}), we find 
\begin{eqnarray}
 &&
 \int d(\delta W)\, \exp\left[
 - \int d\vec{r}\,\mbox{str}\, \left(\delta Q(\vec{r})\right)^{2}\right]
 \nonumber \\
 &=&
 \exp\left[ 2V\mbox{str}\, (\delta T_{0}'-\Lambda\delta T_{0}'\Lambda)\delta T_{0}'
 \right] \nonumber \\
 && 
 \times \left\{
 1+\int d\vec{r}\, \mbox{str}\, \left(\delta T_{0}'-\Lambda\delta T_{0}'\Lambda\right)
 \left( W^{2}\delta T_{0}'-W\delta T_{0}'W\right) \right. \nonumber \\
 &&
 +\frac{1}{2}\left[
 \int d\vec{r}\, \mbox{str}\, \left(\delta T_{0}'-\Lambda\delta T_{0}'\Lambda\right)
 \left( W^{2}\delta T_{0}'-W\delta T_{0}'W\right) \right]^{2} 
 +\frac{1}{64V^{2}} \left(\int d\vec{r}\,\mbox{str}\, W^{2}\right)^{2}
  \nonumber \\
 &&
 +\frac{1}{8V}\int d\vec{r}\,d\vec{r'}\, \mbox{str}\, 
 \left( \delta T_{0}'-\Lambda\delta T_{0}'\Lambda \right)
 \left( 2W(\vec{r})^{2}W(\vec{r'})^{2}\delta T_{0}'
 -2W(\vec{r})^{2}W(\vec{r'})\delta T_{0}'W(\vec{r'}) \right. \nonumber \\
 && \left.
 -2W(\vec{r})W(\vec{r'})^{2}\delta T_{0}'W(\vec{r'})
 +2W(\vec{r})^{2}\delta T_{0}'W(\vec{r'})^{2}
 +W(\vec{r})W(\vec{r'})\delta T_{0}'W(\vec{r'})W(\vec{r})\right)
 \nonumber \\
 && \left.
 -\frac{1}{8}\int d\vec{r}\, \mbox{str}\,
 \left(\delta T_{0}'-\Lambda\delta T_{0}'\Lambda \right)
 W^{2}\delta T_{0}'W^{2}
 -\frac{1}{64V} \int d\vec{r}\,\left(\mbox{str}\, W^{2}\right)^{2}
 +O(W^{5}) \right\}. \label{intdW}
\end{eqnarray}
The remaining $\delta T_{0}$ integral can be also calculated with use of the Wick theorem and the contraction rules:
\begin{equation}
 \langle \mbox{str}\, {\cal X} (\delta T_{0}')_{12} {\cal Y} (\delta T_{0}')_{21} 
 \rangle_{\delta T_{0}} 
 = -\frac{1}{8V} \mbox{str}\, {\cal X} \; \mbox{str}\, {\cal Y} ,
\end{equation}
\begin{equation}
 \langle \mbox{str}\, {\cal X} (\delta T_{0}')_{12} \; \mbox{str}\, {\cal Y} (\delta T_{0}')_{21} 
 \rangle_{\delta T_{0}}
 = -\frac{1}{8V} \mbox{str}\, {\cal X} {\cal Y} ,
\end{equation}
where ${\cal X}$ and ${\cal Y}$ are arbitrary supermatrices and
\begin{equation}
 \langle \ldots \rangle_{\delta T_{0}} = 
 \int d(\delta T_{0})\, (\ldots) \exp\left[ 2V\mbox{str}\, 
 (\delta T_{0}'-\Lambda\delta T_{0}'\Lambda)\delta T_{0}' \right] .
\end{equation}
Since the zero mode variables enter into Eq.~(\ref{intdW}) only through 
$(\delta T_{0}')_{12}$ and $(\delta T_{0}')_{12}$, the Jacobian does not contain the zero mode variables. We finally obtain the Jacobian (\ref{Jacob}) from the expression 
\begin{equation}
 \int d(\delta W) d(\delta T_{0}) \exp\left[
 - \int d\vec{r}\,\mbox{str}\, \left(\delta Q(\vec{r})\right)^{2}\right]
 =1+\frac{1}{16 V^{2}}\int d\vec{r} d\vec{r'} 
 \left(\mbox{str}\, W_{12}(\vec{r}) W_{21}(\vec{r'})\right)^{2}+O(W^{5}).
\end{equation}

\section{Block diagonalization}

We can diagonalize $Y = \int d\vec{r}\, \tilde{Q}(\vec{r})$ in the retarded-advanced space as follows. For Eq.~(\ref{BD}), 
\begin{equation}
Y = \sum_{n=0} Y^{(n)},
\end{equation}
\begin{equation}
 X=\sum_{n=0} X^{(2n+1)},\makebox[0.5cm]{} \left\{X^{(2n+1)},\Lambda\right\}=0,
\end{equation}
\begin{equation}
 \hat{Q}=\sum_{n=0} \hat{Q}^{(2n)},\makebox[0.5cm]{} [\hat{Q}^{(2n)},\Lambda]=0,
\end{equation}
where $Y^{(n)}$, $X^{(n)}$, $\hat{Q}^{(n)}$ mean $O(W^{n})$ term for $Y, X, \hat{Q}$ and $\left\{ , \right\}$ stands for an anticommutator.
Then $\hat{Q}$ and $X$ are perturbatively determined and given by 
Eq.~(\ref{BDQ}) and 
\begin{equation}
 X=\frac{1}{8 V}\int d\vec{r}\, W^{3}(\vec{r})+O(W^{5}).
\end{equation}

We can parametlize $e^{X} T_{0}$ in terms of ${\cal T}$ and $R$:
\begin{equation}
 e^{X} T_{0} = {\cal T} R, \label{bdt}
\end{equation}
where ${\cal T}$ and $R$ obey the properties 
\begin{equation}
 {{\cal T}}^{-1}=\Lambda {\cal T} \Lambda,
 \makebox[0.5cm]{} [R,\Lambda]=0, 
\end{equation}
and are given by 
\begin{equation}
 {\cal T}=\left(e^{X} T_{0}^{2} e^{X}\right)^{1/2}=T_{0}+O(W^{3})+O(W^{5}),
 \label{ht0t0}
\end{equation}
\begin{equation}
 R={{\cal T}}^{-1} e^{X} \left(e^{-X} {{\cal T}}^{2} e^{-X} \right)^{1/2}={\bf 1} + O(W^{3})+O(W^{5}).
\end{equation}
From Eq.~(\ref{ht0t0}), 
\begin{equation}
 dT_{0}=\left(1+O(W^{3})+O(W^{5})\right) d{\cal T} \,\, . \label{jbd}
\end{equation}
Substituting the transformation (\ref{BD}) and (\ref{bdt}) in 
Eqs.~(\ref{Senergy}) and (\ref{Ssource}), we find 
\begin{equation}
 \int d\vec{r}\, \mbox{str}\, \Lambda Q(\vec{r}) 
 =\mbox{str}\, \Lambda {{\cal T}}^{-1} \hat{Q} {\cal T} ,
\end{equation}
\begin{equation}
 \left( \int d\vec{r}\, \mbox{str}\, k \Lambda Q(\vec{r}) \right)^{2}
 =\left( \mbox{str}\, k \Lambda {{\cal T}}^{-1} \hat{Q} {\cal T} \right)^{2}
  +O(W^{3})+O(W^{5}). \label{st}
\end{equation}
 In  Eqs.~(\ref{jbd}) and (\ref{st}), 
$O(W^{3})$ terms contribute to $O(W^{6})$ in $R(s)$
 and are negligible.
 Therefore we replace ${\cal T}$ with $T_{0}$.



\begin{references}
\bibitem{dffsv} The supersymmetric $\sigma$ model description is valid under the condition: $\omega \ll 1/\tau$ where $\tau$ is the elastic mean free time (see Ref.~\cite{Ef}). The energy difference $\omega$ considered in the paper is assumed to satisfy this inequality.
\bibitem{s}
Eq.~(\ref{d2sI}) contains $1/s$ and $1/s^2$ factors.
One may wonder what order of $1/g$ should be assigned to 
these factors because they can become arbitrary large when  
$s$ goes to 0.
However KM's result (\ref{KMr}) shows 
these factors combined with all other $s$ dependent terms give 
no divergence at $s=0$.
Hence we need not take these factors into account for order estimation.
\bibitem{regulator}
In AA's calculation, technically speaking, they had to introduce an external parameter in order to avoid the divergence of their integral over zero mode 
variables of the Fermion-Fermion block for the novel saddle point. 
However, such an artificial regulator is not necessary in 
our calculation.
\end{references}
\end{document}